# Enhancing diamond color center fluorescence via optimized plasmonic nanorod configuration

András Szenes[1], Balázs Bánhelyi[2], Lóránt Zs. Szabó[1],
Gábor Szabó[1], Tibor Csendes[2], Mária Csete[1,*]

[1]Department of Optics and Quantum Electronics, University of Szeged, Dóm tér 9, Szeged, H-6720, Hungary.
[2]Institute of Informatics, University of Szeged, Árpád tér 2, Szeged, H-6720, Hungary
*mcsete@physx.u-szeged.hu, +36-62-544654

**Abstract**
A novel numerical methodology has been developed, which makes possible to optimize arbitrary emitting dipole and plasmonic nano-resonator configuration with an arbitrary objective function. By selecting quantum efficiency as the objective function that has to be maximized at preselected Purcell factor criteria, optimization of plasmonic nanorod based configurations has been realized to enhance fluorescence of NV and SiV color centers in diamond. Gold and silver nanorod based configurations have been optimized to enhance excitation and emission separately, as well as both processes simultaneously, and the underlying nanophotonical phenomena have been inspected comparatively. It has been shown that considerable excitation enhancement is achieved by silver nanorods, while nanorods made of both metals are appropriate to enhance emission. More significant improvement can be achieved via silver nanorods at both wavelengths of both color centers. It has been proven that theoretical limits originating from metal dielectric properties can be approached by simultaneous optimization, which results in configurations determined by preferences corresponding to the emission. Larger emission enhancement is achieved via both metals in case of SiV center compared to the NV center. Gold and silver nanorod based configurations making possible to improve SiV centers quantum efficiency by factors of 1.18 and 5.25 are proposed, which have potential applications in quantum information processing.

**Acknowledgements**
The research was supported by the National Research, Development and Innovation Office-NKFIH through project "Optimized nanoplasmonics" K116362. Mária Csete acknowledges that the project was supported by the János Bolyai Research Scholarship of the Hungarian Academy of Sciences. The authors would like to thank helpful discussions with Professor Niek van Hulst and Professor Lukas Novotny concerning the plasmonic enhancement of fluorescence emission as well as Professor Fedor Jelezko and Professor Ádám Gali regarding the intrinsic quantum efficiency of color centers.

The final publication is available at Springer via http://dx.doi.org/10.1007/s11468-016-0384-1

# 1. Introduction

Spontaneous photon emission from an excited atom or molecule can be tailored by engineering its nanophotonic environment [1]. Already the earliest studies revealed that an emitter in proximity of a dielectric particle exhibits a radiative rate and a quantum efficiency modified with respect to one located in a homogeneous environment [2]. Tiny metal spheroids can be treated as optical antennas, with a resonant frequency depending on their aspect ratio [3]. The interaction of different metallic objects with dipolar emitters has been thoroughly investigated. The reason of quenching occurring at small (~1 nm) distances has been first identified as the interference of out-of-phase dipoles on the emitter and the metal nanoparticle [4]. The strong ~$10^3$-$10^4$-times relaxation rate enhancement was explained by efficient coupling to plasmons [5]. It has been shown that low-order resonances on metallic spheres are capable of promoting the radiative losses as well [6]. Enhancement of absorption and emission were compared in coupled fluorescent molecule and nanoparticle systems [7]. The ~10-times increased radiative rate in metal nano-apertures was attributed to the coupling between molecules and plasmons [8].

The effects of coupled configuration parameters, including the geometry of the nano-objects, and the position and orientation of the emitters, were thoroughly studied. Significant spontaneous emission enhancement has been shown in case of dipoles perpendicular to semi-infinite media and grating-grooves (~10-times), along the axes of dimers (~$10^5$-times) and perpendicularly to axes of rings ($10^2$-times) [9]. It was shown that the fluorescence enhancement is red-shifted with respect to the plasmon resonance [10]. The continuous transition from fluorescence quenching to enhancement was also demonstrated [11].

It was analyzed, how the decay rate enhancement is determined by the different distance dependence of the radiative and non-radiative rates [12]. It has been also described in the literature that limits caused by resistive losses manifest themselves in at least 50% of plasmon energy inside the metal, moreover it was stated that this amount of energy depends on the dielectric functions but not on the specific nanostucture at a given resonance frequency [13]. In contrast to the predicted 50% limit, larger, e.g. 40-53% and 59% apparent quantum efficiency has been achieved for emitters coupled to dimer-antennas and dimer-antenna-arrays, respectively [14]. The possibility to control the angular emission through intermediate resonant plasmonic antennas was also demonstrated [15].

The existence of finite optimal sphere diameter promoting maximal quantum efficiency has been proven [16]. Moreover, the possibility of radiative rate enhancement in narrow spectral intervals via quadrupolar modes arising in optically denser environment far from the interband transition has been also analyzed in case of different metallic materials. It was shown that the radiative / nonradiative rates exhibit a sphere size and distance dependence according to their origin in dipolar and multipolar volume / interface plasmonic modes [17, 18].

Surprisingly, there are only a few examples of optimized configurations in the literature, and typically the excitation and emission events were independently tailored. Larger / smaller enhancement factor has been achieved via thinner / thicker silica shell in case of dyes having an excitation / emission wavelength close to the dipolar resonance on a gold sphere [19].

It was shown that high fluorescence rate enhancement (~$10^4$) as well as high (>0.5) antenna efficiency can be reached within the nanorod's longitudinal resonance band [20]. Simultaneous enhancement of the excitation and emission has been performed in case of dye molecular emitters via transversal and longitudinal localized surface plasmon resonances (LSPR) of nanorods [21]. Recently the material limits of scattering and absorption have been uncovered, and nanorod based systems approaching these limits have been optimized for plane wave illumination. However, novel configurations fundamentally different from usual antenna-like objects were proposed to improve dipolar emitters [22]. A novel interesting approach is combination of optically thin metal films and nano-scatterers, which made possible to achieve $10^4$ Purcell factor with high (>0.5) quantum yield by sweeping the geometrical parameters [23].

Plasmonic enhancement of single molecule fluorescence has been demonstrated during the earliest studies [24]. Single-photon sources possess limited quantum efficiency, therefore, application of plasmon enhanced light emission seems to be a promising approach to improve their brightness. It has been demonstrated that emission from single-photon sources can be efficiently directed to SPPs guided in nanowires and tips [25].

Development of novel diamond based solid-state single-photon sources (SPS) is an outstanding challenge in quantum information processing (QIP), the practical requirements are stable photon generation carrying quantum information via entangled states and efficient extraction of photons [26-32]. Enhanced emission and lifetime engineering have been described in the literature for nitrogen vacancy (NV) centers embedded into appropriately designed diamond nanowires [33] and solid-state cavities [32]. In diamond entanglement occurs between the nuclear and electron spin (solid state qubit) and polarization of the photon (flying qubit) states. Several examples prove in the literature that transfer of quantum information encoded into entangled states is possible in case of plasmon enhanced SPS, e.g., polarization entanglement is preserved during photon-plasmon conversions [34, 35]. Accordingly, integration of plasmonic structures into diamond QIP devices has been performed, and the directivity of emission has been improved via plasmonic aperture-arrays [36-39]. By using plasmonic resonators 75-times enhancement has been achieved due to the small mode volume and large quality factor [40].

Diamond color centers have unique spectroscopic properties, in case of negatively charged NV center, 2.7% of the total emission occurs close to the zero-phonon line at 637 nm with ~90-100% intrinsic quantum efficiency [41]. SiV centers are particularly interesting, as they possess 70% of fluorescence concentrated at 738 nm, and emit light with 83% spectral overlap in HPHT diamond, which makes possible to produce indistinguishable photons with good efficiency [42-44]. However, SiV centers have much lower ~10-15% intrinsic quantum efficiency, moreover, it has been proven that photoluminescence from SiV centers shows a polarization perpendicular to the excitation [44]. According to the latest results in the literature, predesigned plasmonic structures make possible to realize Purcell enhanced spin readout as well as to promote signal transportation with good efficiency [45, 46], however, the existence of an ideal Purcell factor is still a subject of research [45].

In most of the previous studies FDTD method (mainly Lumerical) has been applied to determine the characteristics of coupled emitter - plasmonic nano-object systems via parametric sweeps [19, 20, 37, 40, 48], there is an example of BEM combined with standard optimization methods [22], and only a few examples based on FEM (COMSOL) [39, 46, 47]. This is due to that the built-in analysis groups promote emitter-resonator coupled system characterization via Lumerical, while analogous built-in options are still missing in COMSOL.

The primary purpose of our work was to provide a novel theoretical approach, which makes possible to determine plasmon-emitter configurations possessing the maximal or an arbitrary user-defined quantum efficiency ($QE$) within the limits determined by material dielectric properties; the *Purcell* factor, which is the total decay rate enhancement; as well as the *Purcell·QE* product of these quantities, which is radiative rate enhancement, when there is no quenching without LDOS change and the emitter does not have an intrinsic loss [23, 48]. Secondary purpose was to apply COMSOL and an in-house developed algorithm to optimize the coupled emitter - plasmonic nano-object configurations numerically. The particular subjects of our present study were diamond color center - spheroid coupled systems.

We have selected spheroidal nanorods, since these are historically the primary, and regarding the analytical approaches, the simplest plasmonic objects. Although, there are well known limits in the achievable Purcell factors and quantum efficiencies in case of plane wave illumination, spheroidal nanorods can be tailored to reach the material limits to optical responses [22], therefore they are ideal to demonstrate the capabilities of the novel optimization approach in dipolar emitters improvement. Moreover, it is simple to prepare them e.g. via colloid chemistry. The scientific purpose was to enhance the fluorescence by plasmonically improving both the excitation and emission of NV and SiV color centers in diamond.

**2. Method**

In our present study fluorescent dipole - nanorod configurations capable of maximizing the fluorescence light emission from NV and SiV color centers in diamond were determined by using the RF module of COMSOL Multiphysics. Our purpose was to maximize the (i) excitation and (ii) emission, and finally (iii) both the excitation and emission via coupled color center - nanorod systems. Accordingly, in (i) and (ii) cases single point dipoles oscillating at the wavelength of excitation (532 nm) and emission (650 nm was used to approximate the narrow ZPL of NV, where the fluorescence intensity is similar on the broad spectral peak, and 738 nm was used for SiV) were embedded into diamond dielectric media surrounding small gold and silver nanorods (insets in Fig. 1-3).

To perform (iii) optimization simultaneously at the wavelengths of excitation and emission, in case of NV centers two parallel dipoles corresponding to 532 nm and 650 nm were implemented (insets at top in Fig. 5). In case of SiV, two dipoles oscillating at 532 nm and 738 nm, which are perpendicular to each other, were included into diamond (insets at bottom in Fig. 5), according to the literature [44]. These dipoles and dipole pairs were moved and rotated in the azimuthal plane of the nanorods inside a thin diamond layer having a thickness of 25 nm. According to this, only the $\varphi$ inclination angle was varied during optimization, while the azimuthal orientation was kept constant: $\theta=90°$ (insets in Figure 1-3, 5). The inspected nanorods were qualified by their $a_s$ short and $a_l$ long axes, which were independently tuned. These rods were created as a union of two half-spheres and a cylinder, accordingly the radius of curvature of their caps equals to the half-short-axes of the cylinders. The minimal distance of the dipoles from the nanorods was 2 nm during optimization. The lossy metallic material (gold and silver), and the diamond dielectric medium were specified by tabulated datasets of their wavelength dependent dielectric constants and refractive index, respectively [49]. The diamond coated nanorods were included into air medium, which was bounded by a PML layer, and scattering boundary condition was applied on the outmost surface.

The fluorescence rate enhancement of the dipolar emitter is usually considered according to the literature by analyzing the emission and excitation enhancement relationship [50]:

$$\frac{\gamma^{emission}}{\gamma_0^{emission}} = \frac{QE}{QE_0} \cdot \frac{\gamma^{excitation}}{\gamma_0^{excitation}}, \qquad (1)$$

where the emission enhancement is qualified by the radiative rate enhancement at the wavelength of emission: $\gamma^{emission}/\gamma_0^{emission} = \gamma^{radiative}/\gamma_0^{radiative} = F^{radiative}$, while $QE$ refers to the apparent antenna efficiency of the coupled system and $QE_0$ qualifies the intrinsic efficiency of the emitter in homogeneous environment at the wavelength of emission, respectively.

Generally, the excitation rate depends on the relative orientation of the dipolar emitter with respect to the **E**-field oscillation direction in the exciting light, namely: $\gamma^{excitation}/\gamma_0^{excitation} = |p \cdot E|/|p_0 \cdot E_0|$. In our present study point-like dipoles are used directly to model the color centers, i.e., it is supposed that the excitation **E**-field oscillates parallel to the dipole to be enhanced. According to the reciprocity theorem, the excitation enhancement can be substituted by the radiative rate enhancement at the wavelength of excitation:

$$\left.\frac{\gamma^{radiative}}{\gamma_0^{radiative}}\right|_{emission} = \left.\frac{QE}{QE_0}\right|_{emission} \cdot \left.\frac{\gamma^{radiative}}{\gamma_0^{radiative}}\right|_{excitation} . \quad (2)$$

The complete effect of the plasmonic nano-resonator can be described by considering the modification of the total decay rate. The total decay rate enhancement in close proximity of a plasmonic resonator with respect to the total decay rate in a homogeneous environment can be computed as:

$$F^{total} = \frac{\gamma^{radiative}_{uncoupled\,emitter} + \gamma^{radiative}_{coupled\,resonator} + \gamma^{non-radiative}_{emitter\,intrinsic} + \gamma^{non-radiative}_{coupled\,resonator} + \gamma^{non-radiative}_{quenching}}{\gamma^{radiative}_{uncoupled\,emitter,0} + \gamma^{non-radiative}_{emitter\,intinsic,0}} . \quad (3)$$

The $QE_0 = \gamma^{radiative}_{uncoupled\,emitter,0}/(\gamma^{radiative}_{uncoupled\,emitter,0} + \gamma^{non-radiative}_{emitter\,intrinsic,0})$ intrinsic quantum efficiency related to the emitter is taken into account generally, however, the intrinsic loss can be supposed to be equal in the absence and presence of plasmonic nanorods: $\gamma^{non-radiative}_{emitter\,intrinsic} = \gamma^{non-radiative}_{emitter\,intrinsic,0}$. In case of emitters having unity $QE_0$ and in absence of quenching resulting in $\gamma^{non-radiative}_{quenching} = 0$, the $F^{total}$ total decay rate enhancement simplifies to the quotient of the sum of radiative and non-radiative rates and the radiative rate in homogeneous environment, which is nominated as the *Purcell* factor [23, 47, 48]:

$$Purcell = F^{total}\Big|_{\gamma^{non-radiative}_{emitter\,intrinsic}=0,\,\gamma^{non-radiative}_{quenching}=0} . \quad (3a)$$

We have selected the *Purcell* factor as one of the parameters appropriate to qualify the configurations, and at each specific wavelength of interest, the *Purcell* factor was computed according to the literature, by calculating the quotient of power outflows from the dipole in absence and presence of the nanorod [51]:

$$Purcell\big|_{read-out} = \frac{P^{radiative} + P^{non-radiative}}{P_0^{radiative}} . \quad (3b)$$

Here the $P^{radiative} + P^{non-radiative}$ total emitted power was determined by reading out the power flow from a tiny imaginary sphere surrounding the dipole, while the $P^{radiative}$ was calculated from the power-outflow through a spherical region including the coupled system, just below a PML layer. The $P^{non-radiative}$ heat-loss was computed based on the resistive heating inside the nanorod as well. The coupled system's quantum efficiency is as follows:

$$QE = \frac{\gamma^{radiative}_{uncoupled\,emitter} + \gamma^{radiative}_{coupled\,resonator}}{\gamma^{radiative}_{uncoupled\,emitter} + \gamma^{radiative}_{coupled\,resonator} + \gamma^{non-radiative}_{emitter\,intrinsic} + \gamma^{non-radiative}_{coupled\,resonator} + \gamma^{non-radiative}_{quenching}} . \quad (4)$$

The $QE$ simplifies to the quotient of the sum of radiative rates and the sum of the radiative and non-radiative rates, when $\gamma^{non-radiative}_{emitter\,intrinsic}$ intrinsic decay rate of the emitter and the $\gamma^{non-radiative}_{quenching}$ quenching related rate are zero:

$$QE\Big|_{\gamma^{non-radiative}_{emitter\,intrinsic}=0,\,\gamma^{non-radiative}_{quenching}=0} = \frac{\gamma^{radiative}_{uncoupled\,emitter} + \gamma^{radiative}_{coupled\,resonator}}{\gamma^{radiative}_{uncoupled\,emitter} + \gamma^{radiative}_{coupled\,resonator} + \gamma^{non-radiative}_{coupled\,resonator}} \quad (4a)$$

The quantum efficiency was read-out accordingly as follows:

$$QE\big|_{read-out} = \frac{P^{radiative}}{P^{radiative} + P^{non-radiative}} . \quad (4b)$$

The product of the total decay rate enhancement and the $QE$ was determined to rank the optimized systems:

$$F^{total} \cdot QE = \frac{\gamma_{uncoupled\,emitter}^{radiative} + \gamma_{coupled\,resonator}^{radiative}}{\gamma_{uncoupled\,emitter,0}^{radiative} + \gamma_{dipole\,intrinsic,0}^{non\text{-}radiative}}. \quad (5)$$

This relationship can be rewritten in a more informative form by dividing each quantity in Eq. (5) by the radiative decay rate in absence of the nanoparticle, which shows that the product simplifies to the $F^{radiative}$ radiative rate enhancement, when the $\gamma_{emitter\,intrinsic,0}^{non-radiative}$ intrinsic loss is zero:

$$F^{total} \cdot QE = \frac{\frac{\gamma_{uncoupled\,emitter}^{radiative} + \gamma_{coupled\,resonator}^{radiative}}{\gamma_{uncoupled\,emitter,0}^{radiative}}}{1 + \frac{1-QE_0}{QE_0}} = \frac{F^{radiative}}{1 + \frac{1-QE_0}{QE_0}}. \quad (5a)$$

The radiative rate enhancement can be read out based on Eq. 3, 3a, 3b, 4, 4a, 4b as follows:

$$F^{radiative}\big|_{read\text{-}out} = Purcell \cdot QE\big|_{read\text{-}out} = \frac{P^{radiative}}{P_0^{radiative}} \quad (5b)$$

The *QE* of the coupled system in case in non-unity intrinsic $QE_0$ can be computed with a rescaling:

$$QE^{corrected} = \frac{P^{radiative}}{P^{radiative} + P^{non\text{-}radiative} + \frac{1-QE_0}{QE_0}}. \quad (6)$$

Both Eqs. (5) and (5a) indicate that in case of unity $QE_0$ of the emitters and in absence of quenching the *Purcell·QE* product equals to the radiative rate enhancement specified in Equations (1) and (2). However, these relationships indicate that the $F^{radiative}$ radiative rate enhancement at the excitation wavelength and the radiative rate enhancement as well as the *QE* that are reachable at the emission wavelength, are interdependent, accordingly we expect barriers in case of simultaneous optimization.

First, we have performed a conditional optimization of the *Purcell* and *Purcell*$_{excitation}$·*Purcell*$_{emission}$ factors (details can be found in ref. [52, 53]). From the point of view of applications, it is more straightforward to realize a conditional optimization using the *QE* and *QE*$_{excitation}$·*QE*$_{emission}$ objective functions by setting a criterion regarding the *Purcell* factor that have to be met independently at the excitation and emission wavelength in cases of (i) and (ii), and by setting parallel criteria on *Purcell*$_{excitation}$ and *Purcell*$_{emission}$ factors that have to be met simultaneously at these wavelengths in (iii) case. These objective functions are in accordance with the intuitive expectation that configurations optimal to enhance excitation and emission both correspond to those constellations, which ensure to reach a desired level of *Purcell* factor with the maximal *QE* achievable in the coupled system. Selection of the same *QE* objective function at the excitation wavelength can be explained with that any dipole can be most efficiently excited, when the smallest fraction of injected energy is absorbed inside the nanorod. Accordingly, evaluation of each configuration was performed by computing the *Purcell* factor, the *QE* quantum efficiency and the product of them. For each *Purcell* criterion / *Purcell*$_{excitation}$ and *Purcell*$_{emission}$ criteria the configuration resulting in the highest *QE* / *QE*$_{excitation}$·*QE*$_{emission}$ has been determined, and the optimal points as a function of the distance parameter and *Purcell* criterion / *Purcell*$_{excitation}$ and *Purcell*$_{emission}$ criteria are visualized in Fig. 1, 2, 3 / Fig. 5.

Algorithms built into COMSOL are not suitable to perform optimization for two wavelengths simultaneously, therefore application of an external algorithm has been necessary. The dipole-nanorod configuration optimization has been performed by implementing the GLOBAL algorithm via LiveLink for Matlab, which robustness has been proven in different application areas [54-56]. GLOBAL is suitable to find the global extrema within few objective function evaluation steps by mapping the local extrema first. Particular advantage is that GLOBAL provides automatically a robust solution, namely a point in a parameter-space that can slightly be changed without dramatic differences in the objective function value, which feature is important in experimental realization of the obtained nanophotonical designs. GLOBAl has been successfully tested up to 15 parameters, in our present optimization procedure the short and long nanorod axes dimension ($a_s$, $a_l$), as well as the inclination angle ($\varphi$) specifying the dipole orientation and the ($x$, $y$, at $z=0$) dipole position were the varied parameters.

We have determined and analyzed the optimal configurations for the (i) 532 nm excitation wavelength first, then (ii) for the emission wavelengths of NV and SiV centers, finally (iii) the configurations optimized simultaneously at the excitation and emission wavelengths were studied. The initial and the final points of optimizations resulting in optimal configurations are indicated in Figs. 1-3 and Fig. 5, which show that the *Purcell* factor values taken on approximate the initial criterions in case of (i) and (ii) optimizations, while the (iii) simultaneous optimizations resulted in *Purcell* factors, which differ more significantly from the initial points.

Although, the *Purcell* factor criteria were varied in wider intervals, tendencies throughout *Purcell*$_{crit}$=1 were analyzed in all cases, since this criterion corresponds to the limit of total decay rate enhancement.

Configurations exhibiting the highest *QE·Purcell* and *(QE·Purcell)*$^2$ product were selected for detailed wavelength dependent studies performed for single wavelength of excitation and emission and for both wavelengths simultaneously. The *Purcell* factor, *QE* and *Purcell·QE* was determined as a function of wavelength by sweeping the frequency of the dipole in [400 nm, 900 nm] interval with ~1 nm resolution (Fig. 1-3, 5, 6). The time averaged power flow and **E**-field distribution, as well as the accompanying charge distribution were inspected at the excitation and emission to understand the underlying physics (Fig. 4, 8). First, all quantities were determined at the maxima by supposing unity *QE*$_0$ for the color centers, then the non-unity *QE*$_0$ values were taken into account to determine the realistic *QE* according to Eq. (6).

The comparative study of the optimized configurations based on gold and silver nanorods was realized. The comparison of configurations determined by (i) and (ii) optimizations for single wavelength, and (iii) optimization for double wavelengths was performed as well. In addition to this, the comparison of results originating from (iii) optimization of NV and SiV configurations is presented as well. All relevant numerical results of the optimizations are collected in comparative electronic Online Resource, Table 1 and Table 2.

**3. Results**

***3.1 Gold and silver nanorod configurations optimized for single wavelength***

The optimizations performed for single wavelength have shown that the *QE* decreases, when the value of *Purcell* factor criterion increases, and the highest achievable value of the *QE* quantity as well as of the *Purcell·QE* product depends on the type of metallic material and the wavelength, where the optimization has been performed (Figures 1-3/a, b). The achieved results show that by increasing the wavelength the loss can be decreased according to the decreasing amount of plasmon energy in metal [13], and to the decreasing role of interband transitions [16, 57], while larger radiative rate enhancement is achievable according to the material limits to optical responses manifesting themselves in increasing per-volume normalized scattering cross-section [22]. All material limits are better at the same wavelength in case of silver, therefore larger emission enhancement with better efficiency is expected in case of silver nanorod based coupled systems.

***3.1.1 Gold and silver nanorod configurations optimized for single wavelength of excitation***

The achievable *QE* rapidly increases in the same [0, 10$^2$] interval by decreasing the criterion regarding the *Purcell* factor through unity, and the corresponding emitter distance increases to 21.59 nm and 22.70 nm in case of gold and silver nanorod-based configurations, respectively (Fig. 1a, b). There is a cut-off in case of both materials at 10$^4$ *Purcell* factor, indicating that higher total decay rate enhancements are not achievable.

In the optimal gold nanorod based configuration the almost unity 0.96 *Purcell·QE* product is reached at 3006.18 *Purcell* factor, while the extremely low 0.03% *QE* indicates that significant amount of light is absorbed inside the metal spheroid (Online Resource, Table 1, top). The lower than unity *Purcell·QE* product shows that no radiative rate enhancement is achievable in the gold nanorod based configuration at 532 nm. The *Purcell* factor and the *QE* exhibit global maxima at wavelengths slightly and significantly larger than 532 nm, respectively. The wavelength dependency of the *Purcell·QE* product shows the global maximum also at a wavelength larger than 532 nm (Fig. 1c, d). The inappropriate position of the maxima and the complete wavelength dependency of the *Purcell* factor and *QE* indicate that it is hard to design a gold nanorod capable of enhancing NV or SiV excitation at 532 nm, which is close to the gold interband transitions [16, 57].

Two orders of magnitude larger 65.21 *Purcell·QE* product has been reached in silver nanorod based configuration at 1076.43 *Purcell* factor, while the 6.06% *QE* indicates that smaller amount of light is absorbed inside the silver nanorod (Online Resource, Table 1, top). The *Purcell·QE* product is larger than unity, indicating significant radiative rate enhancement in the silver nanorod based configuration at 532 nm. The *Purcell* factor and *QE* shows a local and a global maximum at the wavelength of excitation. As a result, the wavelength dependency of the *Purcell·QE* product shows the global maximum almost exactly at 532 nm (Fig. 1c, d). These tendencies indicate that it is more realistic to design a configuration based on silver nanorod to enhance NV and SiV excitation at 532 nm.

In comparison, the *Purcell·QE* is 68.12-times larger, which originates from a reduced *Purcell* factor and a two-orders of magnitude larger *QE* in case of silver, indicating that the larger *Purcell·QE* radiative rate enhancement is due to reduced non-radiative loss in silver (Online Resource, Table 1, top). These results are in accordance with that 532 nm is far from silver interband transitions, where smaller fraction of plasmon energy concentrated in the silver nanorod, and the material limits to optical response are significantly better in case of silver [13, 22].

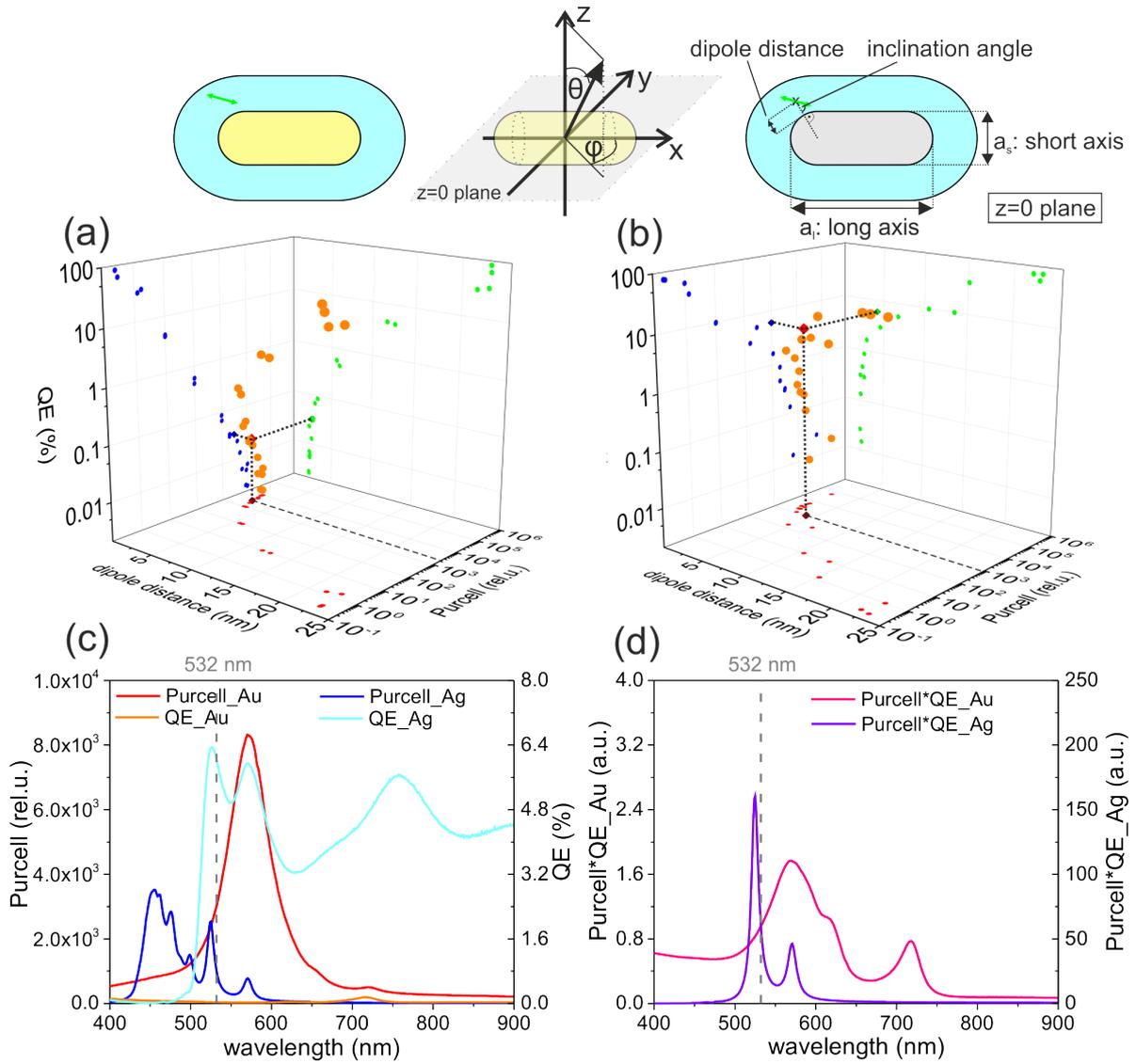

**Fig. 1** (a, b) Interdependence of the *Purcell* factor criterion, *QE* quantum efficiency, corresponding distance for (a) gold and (b) silver nanorod (orange symbols: values taken on, blue / green / red: projections onto 2D parameter planes, corresponding dark (purple / indigo / olive / wine) symbols with dotted-line: global optimum, empty symbol with dashed-line: initial Purcell criterion). Wavelength dependency of the (c) *Purcell* factor and *QE*, and (d) *Purcell·QE* product (gold warm, silver cold colors) in the configuration optimized for single 532 nm of excitation. Insets: schematics of dipole-nanorod configurations

*3.1.2 Gold and silver nanorod configurations optimized for single wavelength of NV emission*

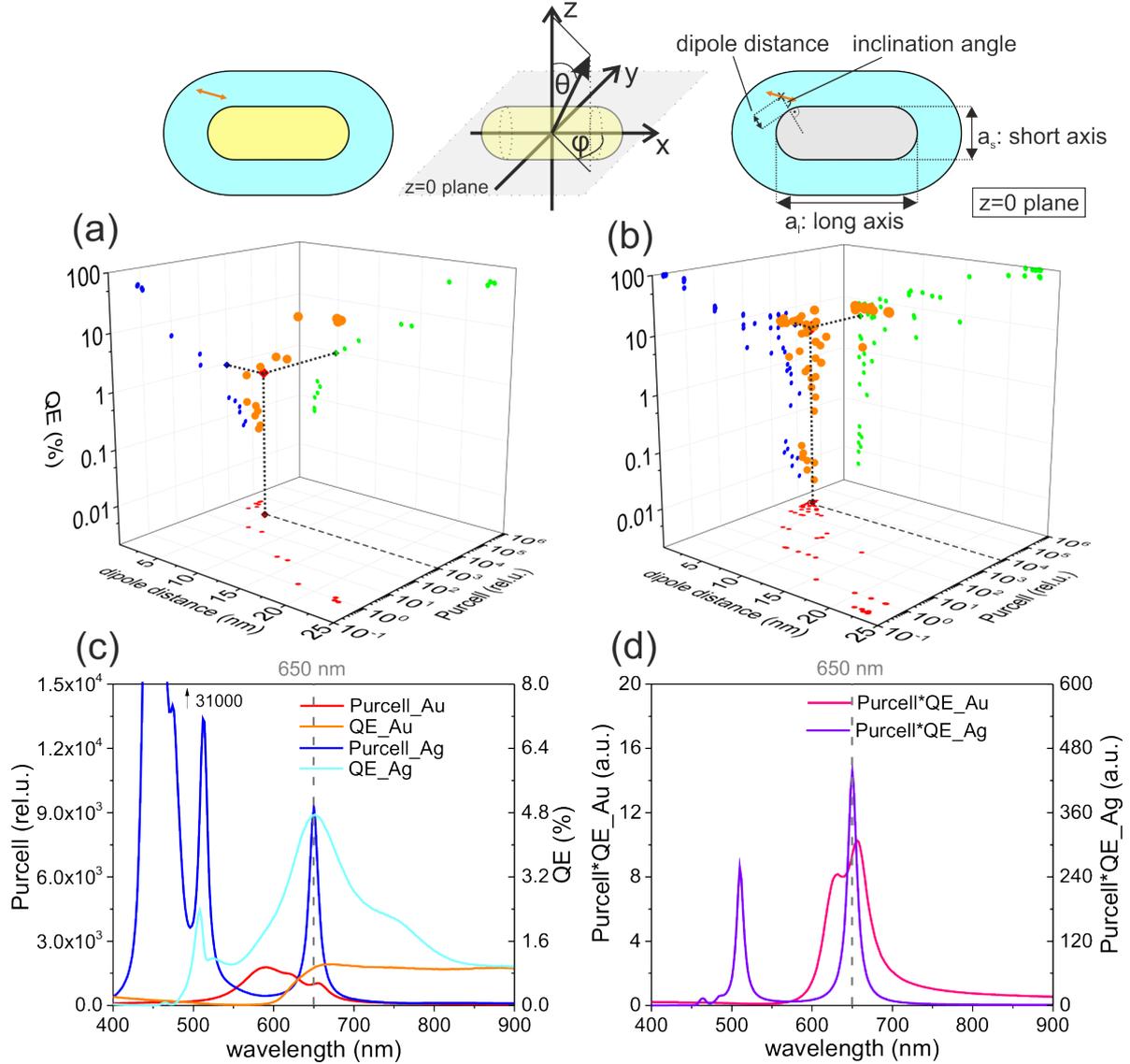

**Fig. 2** (a, b) Interdependence of the *Purcell* factor criterion, *QE* quantum efficiency, corresponding distance for (a) gold and (b) silver nanorod (orange symbol: values taken on, blue / green / red: projections onto 2D parameter planes, corresponding dark (purple / indigo / olive / wine) symbols with dotted-lines: global optimum, empty symbol with dashed-line: initial Purcell criterion). Wavelength dependency of the (c) *Purcell* factor and *QE*, and (d) *Purcell·QE* product (gold warm, silver cold colors) in the configuration optimized for single 650 nm of NV emission. Insets: schematics of dipole-nanorod configurations

At the wavelength of NV emission, the achievable *QE* increases in [$10^{-1}$, $10^{2}$] and [$10^{-2}$, $10^{2}$] intervals by decreasing the criterion regarding the *Purcell* factor through unity more rapidly than at the excitation wavelength (Fig. 2a, b). The corresponding distance increases to 22.17 nm and 19.64 nm in case of gold and silver nanorod, respectively. There is a cut-off in case of both materials at $10^{4}$ *Purcell* factor, indicating that higher total decay rate enhancements are not achievable.

In the optimal gold nanorod based configuration approximately one order of magnitude larger 9.42 *Purcell·QE* product is reached at the wavelength of NV emission, when 1002.01 *Purcell* factor is taken on. The 0.94% *QE* indicates that less significant amount of light is absorbed inside the gold nanorod at 650 nm than at 532 nm (Online Resource, Table 2, top). The *Purcell* factor and *QE* indicates a local and a global maximum at the wavelength of NV emission. As a result, the wavelength dependency of the *Purcell·QE* product shows the global maximum exactly at 650 nm wavelength (Fig. 2c, d).

The possibility of reaching 9.81-times larger *Purcell·QE* via gold nanorod indicates that at the 650 nm emission wavelength, which is far from the interband transitions, the optimized gold nanorod exhibits an LSPR accompanied by significant radiative rate enhancement [16, 57]. The increase in the reachable *QE* and *Purcell·QE* with the wavelength is in accordance with the wavelength dependency of the plasmon energy fraction concentrated in the metal as well as with the wavelength dependency of material limits to optical responses [13, 22]. Our results prove that it is possible to design a gold nanorod capable of improving NV center emission.

In silver nanorod based configuration approximately one order of magnitude larger 438.16 *Purcell·QE* product is reached at the wavelength of NV emission, when 9262.95 *Purcell* factor is taken on, which approximates the highest *Purcell* factor achievable via silver nanorod. The 4.73% *QE* indicates that the fraction of light absorbed inside the silver nanorod is almost equal in silver nanorods optimized at 532 nm and at 650 nm (Online Resource, Table 2, top).

The *Purcell* factor and *QE* indicate a local and a global maximum at the wavelength of NV emission, similarly to gold. As a result, the wavelength dependency of the *Purcell·QE* product shows the global maximum exactly at 650 nm wavelength (Fig. 2c, d). Although, there are huge *Purcell* factor maxima, e.g. at 532 nm as well, the *QE* is very low in this interval, so their product exhibits only a local maximum.

The possibility of reaching 6.72-times higher *Purcell·QE* indicates that at the 650 nm emission wavelength, which is very far from the interband transitions, the optimized silver nanorod exhibits an LSPR accompanied by better radiative rate enhancement compared to that at 532 nm [16, 57]. The significantly higher *Purcell* factor is accompanied by a slightly reduced *QE* with respect to the excitation wavelength, revealing that the optimized configuration does not reach the material related limits to optical responses at 650 nm [13, 22]. Based on these results, it is possible to design a silver nanorod capable of enhancing NV center emission, moreover, the wavelength dependency of optical responses suggests that simultaneous optimization at the excitation and emission wavelength is also probable (Online Resource, Table 2, top, left).

In comparison, in silver nanorod the 46.5-times larger *Purcell·QE* originates from *Purcell* factor and *QE*, which is slightly more / less strongly enhanced with respect to gold, respectively.

### *3.1.3 Gold and silver nanorod configurations optimized for single wavelength of SiV emission*

At the wavelength of SiV emission, the achievable *QE* follows the tendencies observed during the previous optimizations, however increases in [$10^{-2}$, $10^1$] and [$10^{-2}$, $10^2$] intervals by decreasing the criterion regarding the *Purcell* factor through unity more rapidly, than at the excitation and NV emission wavelengths (Fig. 3a, b). The corresponding distance increases to 19.85 nm and 22.97 nm in case of gold and silver nanorod, respectively. There is again a cut-off in both materials at $10^4$ *Purcell* factor, indicating that higher decay rates are not achievable.

In gold nanorod based configuration 44.14 *Purcell·QE* product is reached at the wavelength of SiV emission, when 2050.33 *Purcell* factor is taken on. The 2.15% *QE* indicates that the non-radiative loss inside the nanorod is the smallest at 738 nm (Online Resource, Table 2, top, right). The *Purcell* factor and *QE* indicate again a local and a global maximum at the wavelength of SiV emission. As a result, the wavelength dependency of the *Purcell·QE* product shows the global maximum exactly at 738 nm (Fig. 3c, d).

The *Purcell·QE* is 45.98-times and 4.69-times increased with respect to those at the excitation and NV emission wavelength. The possibility of reaching further increased *Purcell·QE* product is in accordance with that better radiative rate enhancement is achievable at smaller frequencies due to the wavelength dependency of material related limits to optical responses [13, 16, 22, 57]. As a result, designing a gold nanorod based configuration capable of enhancing SiV center emission is possible and the enhancement is more significant, than it can be achieved in case of NV center.

In silver nanorod based configuration, 1548.08 *Purcell·QE* product is reached at the wavelength of SiV emission, when 27720.49 *Purcell* factor is taken on. The 5.58% *QE* is slightly larger than the value reached at 650 nm in silver nanorod based configuration. Both the *Purcell* factor and *QE* indicate a global maximum at the wavelength of SiV emission, which proves that a silver nanorod can be more efficiently optimized than a gold nanorod. There are local *Purcell* factor and *QE* maxima slightly before 532 nm as well, suggesting that optimization simultaneously at the wavelengths of excitation and emission is possible. The wavelength dependency of the *Purcell·QE* product shows the global maximum exactly at 738 nm wavelength, and there is a small local maximum slightly before 532 nm (Fig. 3c, d).

The *Purcell·QE* is 23.74- and 3.53-times larger than at 532 nm and 650 nm, indicating better radiative rate enhancement very far from the interband transitions of silver [16, 57]. The possibility of enhancing *QE* with respect to NV emission is in accordance with that the non-radiative losses decrease due to the smaller plasmon energy fraction in metal, when the wavelength is increased [13]. Huge *Purcell* factor can be reached along with a high *QE*, which results in the most significant radiative rate enhancement in the silver nanorod based system optimized at 738 nm wavelength, in accordance with the wavelength dependency of material related limits to optical responses [22].

In comparison, the 35.07-times larger *Purcell·QE* originates from significantly and slightly larger *Purcell* and *QE* with respect to gold (Online Resource, Table 2, top, right).

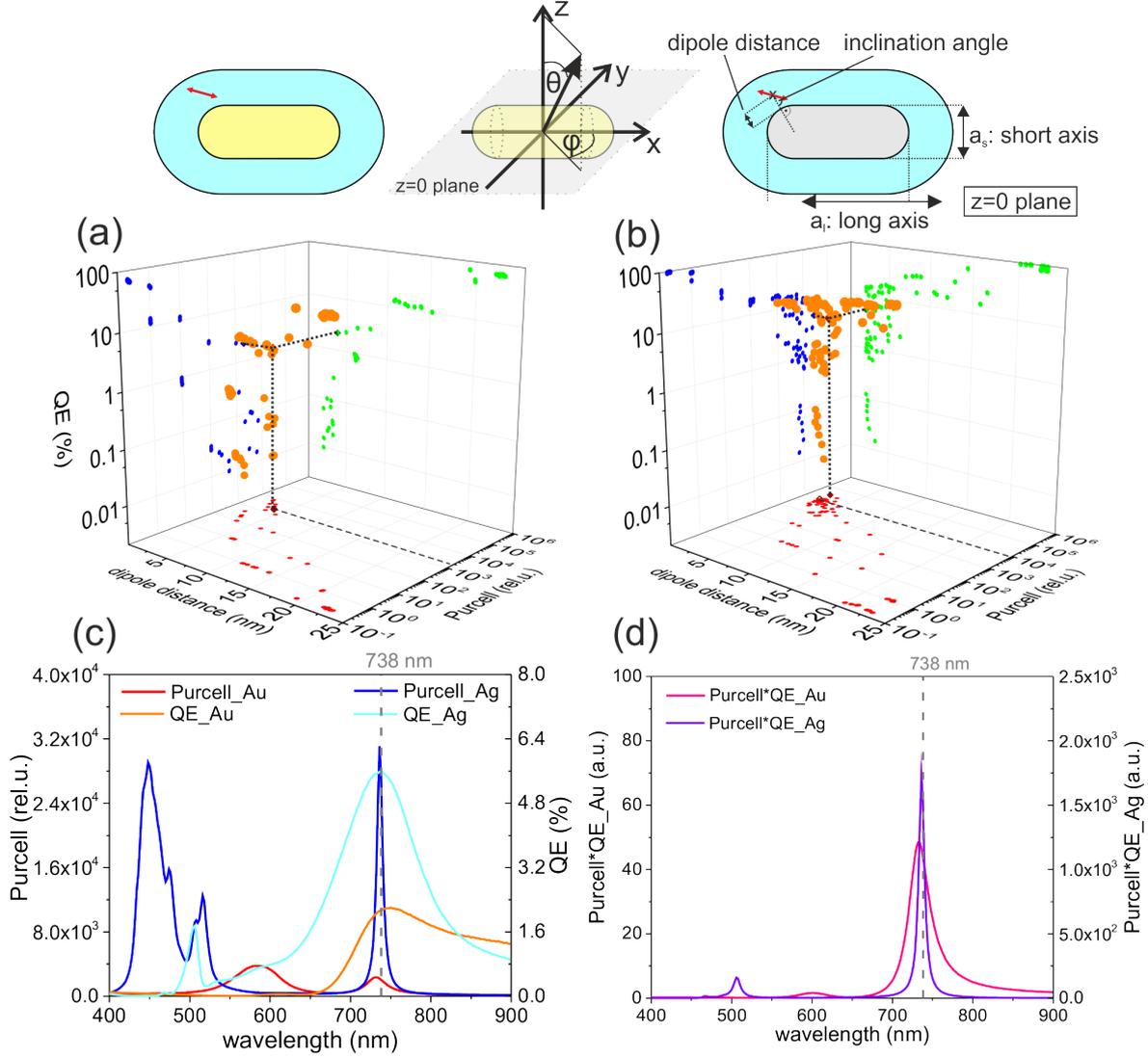

**Fig. 3** (a, b) Interdependence of the *Purcell* factor criterion, *QE* quantum efficiency, corresponding distance for (a) gold and (b) silver nanorod (orange symbol: values taken on, blue / green / red: projections onto 2D parameter planes, corresponding dark (purple / indigo / olive / wine) symbols with dotted-lines: global optimum, empty symbol with dashed-line: initial Purcell criterion) Wavelength dependency of the (c) *Purcell* factor and *QE*, and (d) *Purcell·QE* product (gold warm, silver cold colors) in the configuration optimized for single 738 nm of SiV emission. Insets: schematics of dipole-nanorod configurations

### 3.1.4 Comparative study of near-field and charge distributions in configurations optimized for excitation and emission wavelengths

In configurations optimized for 532 nm wavelength of excitation, a very elongated gold nanorod with 4.40 aspect ratio is the optimal one, while the 1.15 aspect ratio indicates that the silver nanorod is almost spherical. The dipole is located at 2.34 nm from the gold nanorod, while its inclination angle is 74.09°. In case of silver nanorod the distance is 4.16 nm and the 7.97° inclination angle reveals that the dipole is almost perpendicular to the surface.

Caused by the small distance and large inclination angle, a localized dipolar surface mode is observable on the gold nanorod, while the silver nanorod shows dipolar volume distribution. Accordingly, the **E**-field is enhanced locally in proximity of the point dipole and the power flow is trapped at the gold nanorod, while on the silver nanorod more dipolar **E**-field enhancement is observable, which is accompanied by a power flow emanating from the coupled system. These results demonstrate that the localized surface modes cause large non-radiative losses in gold, in accordance with the literature about that a non-radiative decay channel process is associated with coupling to the interface plasmon mode [17]. In contrast, the volume dipolar mode is capable of enhancing the radiative rate in silver, according to that radiative decay processes involve a transfer to a dipolar plasmon mode (Online Resource, Table 1, top) [17].

In configurations optimized for NV center emission almost spherical gold nanorod with 1.12 aspect ratio is optimal, while the silver nanorod with 1.47 aspect ratio is slightly more elongated. The dipole is located at 4.78 nm from the gold nanorod and exhibits -8.36° tilting, while the distance is 2.18 nm from the silver nanorod, which is accompanied by 18.91° tilting. Both inclination angles reveal that the dipoles are almost perpendicular to the gold and silver nanorod surface. Both nanorods show dipolar volume charge distribution acting as a radiative decay channel, however, the larger aspect ratio of the silver nanorod promotes larger radiative rate enhancement (Online Resource, Table 2, top-left). Accordingly, larger and more dipolar **E**-field enhancement is observable around the silver nanorod, which is accompanied by stronger power outflow.

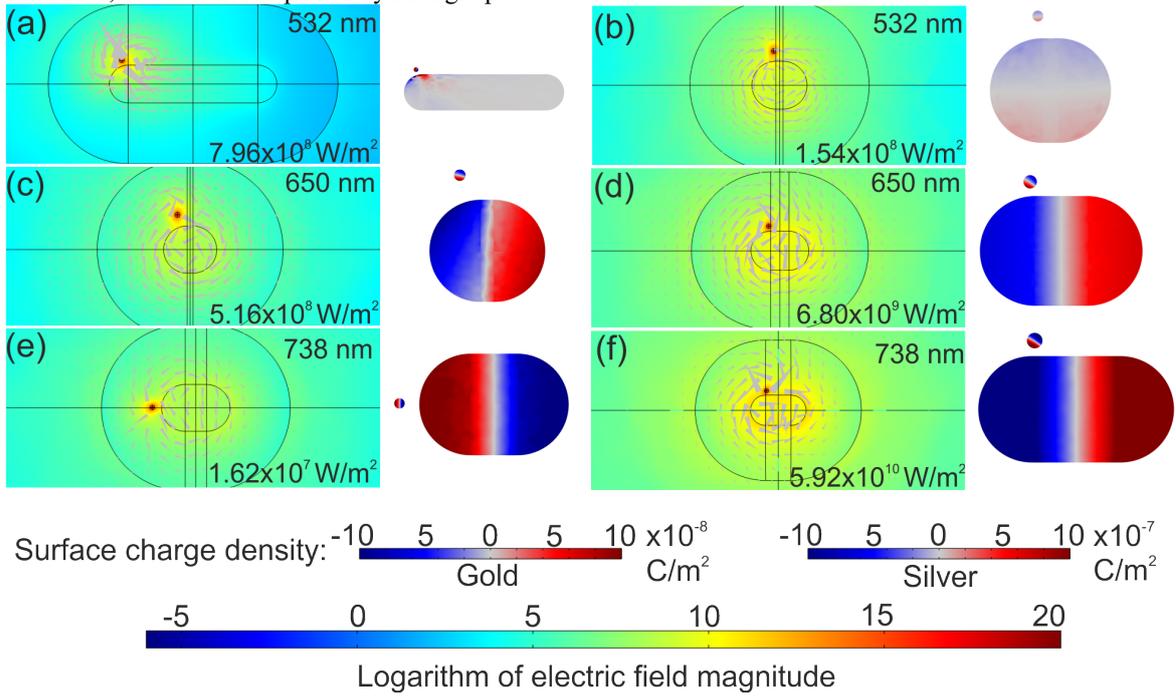

**Fig. 4** Near-field and power flow of optimized configurations at (a, b) excitation, (c, d) NV and (e, f) SiV emission wavelength. Color centers near (a, c, e) gold nanorod and (b, d, f) silver nanorod. Insets: corresponding charge density plots at excitation and emission wavelengths. Note that the arrows indicate the total power flow including the radiated and non-radiated channels in the coupled system.

In configurations optimized for SiV center emission a relatively more spherical gold nanorod with 1.45 aspect ratio is optimal, while the silver nanorod with 1.82 aspect ratio is relatively more elongated. The dipole is located at 3.8 nm from the gold nanorod, while the distance is 2.00 nm from the silver nanorod. The -23.39° inclination angle of the dipole with respect to the normal of gold nanorod surface is almost two times smaller than the 48.10° tilting in case of silver nanorod.

Again, both nanorods show dipolar volume charge distribution, however, the larger aspect ratio of the silver nanorod promotes larger radiative rate enhancement, despite the anomalously large dipole tilting (Online Resource, Table 2, top-right). Accordingly, larger dipolar **E**-field enhancement and much stronger emanating power flow is observable around the silver nanorod, however more antenna-like power outflow is noticeable in case of the gold nanorod due to the on-axes location of the emitter.

*3.2 Comparison of gold and silver rod configurations optimized for wavelengths of excitation and emission simultaneously*

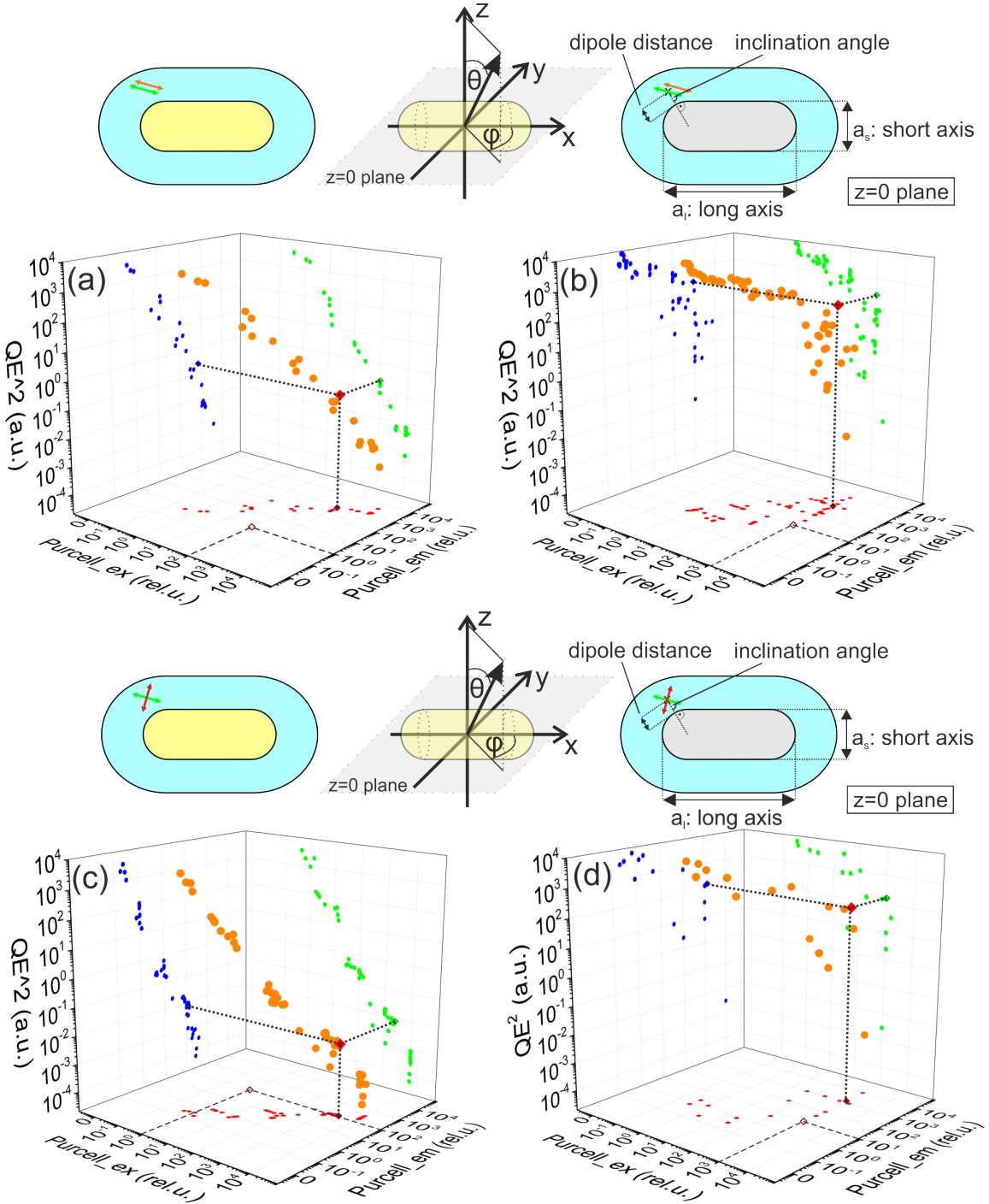

**Fig. 5** Achievable $QE_{\text{excitation}} \cdot QE_{\text{emission}}$ ($QE^2$) in configurations optimized with minimum criterion regarding Purcell factors that have to be met simultaneously at the excitation and emission wavelengths (orange symbols: values taken on, blue / green / red: projections onto 2D parameter planes, corresponding dark (purple / indigo / olive / wine) symbols with dotted-lines: global optimum, empty symbol with dashed-lines: initial Purcell criterion). (a) NV center and gold nanorod, (b) NV center and silver nanorod, (c) SiV center and gold nanorod, (d) SiV center and silver nanorod. Insets: schematics of dipole-nanorod configurations

During optimization performed to maximize the $QE_{excitation} \cdot QE_{emission}$ quantity, the criteria regarding the minimal $Purcell_{excitation}$ and $Purcell_{emission}$ factors that have to be parallel met at the wavelengths of excitation and emission have been stepped in smaller [1, $10^4$] interval. Based on these results one can conclude that the achievable $QE_{excitation} \cdot QE_{emission}$ product values increase by decreasing either of $Purcell$ criteria for both metals. The increase is more rapid and significant, furthermore, it indicates a more well-defined wavelength dependency in case of gold nanorods. However, the achievable $QE_{excitation} \cdot QE_{emission}$ product values are slightly smaller in case of gold nanorod (Fig. 5a-d).

In case of gold nanorod based configuration optimization at 532 nm and 650 nm NV center wavelengths simultaneously, the $QE_{excitation} \cdot QE_{emission}$ product increases monotonously in [$10^{-3}$, $10^3$] interval by decreasing the $Purcell$ criteria through unity, and the increase exhibits similar rate as a function of criteria at the wavelength of excitation and emission (Fig. 5a). In contrast, when the optimization of gold nanorod based configuration is performed at 532 nm and 738 nm SiV center excitation and emission wavelengths simultaneously, the $QE_{excitation} \cdot QE_{emission}$ product increases monotonously in [$10^{-4}$, $10^2$] interval with one order of magnitude smaller bounds. Moreover, the increase is less rapid at the excitation wavelength, namely the smallest $10^{-4}$ values are reached at $10^4$ $Purcell$ criterion at the excitation wavelength, while same values are already taken in [$10^2$, $10^3$] $Purcell$ criterion interval at the emission wavelength (Fig. 5c). A cut-off appears for both gold nanorod based configuration optimizations at ~$10^4$ and ~$10^3$ criterion regarding the minimal $Purcell$ factor at the excitation and emission wavelengths for both of NV and SiV center excitation and emission.

In case of silver nanorod based configuration optimization at 532 nm and 650 nm NV center as well as at 532 nm and 738 nm SiV center wavelengths simultaneously, the $QE_{excitation} \cdot QE_{emission}$ product increases less monotonously in smaller [$10^{-2}$, $10^3$-$10^4$] interval by decreasing the $Purcell$ criterion through unity, and the increase exhibits a similar rate at the wavelength of excitation and emission (Fig. 5b, d). A cut-off appears for both silver nanorod based configuration optimizations at ~$10^3$ and ~$10^4$ criterion regarding the minimal $Purcell$ factor at the excitation and emission wavelengths for both of NV and SiV center excitation and emission.

By comparing the nanorods made of different metallic materials, one can conclude that 6 / 5 orders of magnitude modification occurs inside the inspected $Purcell$ criterion intervals in the achievable $QE_{excitation} \cdot QE_{emission}$ product in case of gold and silver, respectively. Both bounds of the $QE_{excitation} \cdot QE_{emission}$ product intervals are lower in case of SiV than in case of NV, when the centers are in proximity of gold. In case of silver nanorods the achievable values are in similar $QE_{excitation} \cdot QE_{emission}$ interval for both color centers.

*3.2.1 Comparative study of gold and silver rod configurations optimized simultaneously at excitation and emission wavelength of NV*

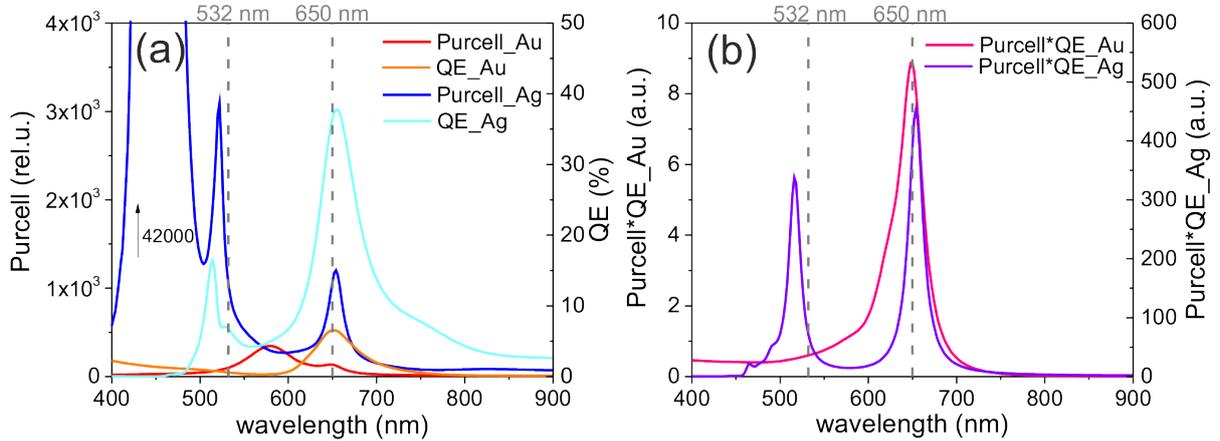

**Fig. 6** Wavelength dependency of the (a) *Purcell* factor and *QE* quantum efficiency, (b) *Purcell·QE* product, in configurations based on gold and silver nanorod, optimized for excitation and emission wavelength of NV center

In gold nanorod based configuration the highest 5.37 $Purcell_{excitation} \cdot Purcell_{emission} \cdot QE_{excitation} \cdot QE_{emission}$ ($Purcell \cdot QE$)$^2$ quantity with 0.6 and 8.92 $Purcell \cdot QE$ products is reached, when 100.35 and 134.77 $Purcell$ factors are taken on. These $Purcell$ factors are accompanied by 0.60% and 6.62% $QE$ values, showing that the non-radiative loss is approximately with an order of magnitude smaller at the wavelength of emission (Online Resource, Table 1, 2 middle, left). The $Purcell \cdot QE$ is less/larger than unity at 532 nm / 650 nm, which shows again no/considerable radiative rate enhancement.

The *Purcell* factor and *QE* shows a local and a global maximum at the 650 nm wavelength of emission, however, neither of them shows a maximum at 532 nm (Fig. 6a). On the contrary, there is a global maximum on the *Purcell* factor in between 532 nm and 650 nm. Similarly, the wavelength dependency of the *Purcell·QE* product shows the global maximum exactly at 650 nm, while does not exhibit a local maximum at 532 nm (Fig. 6b). The tendencies indicate that the net fluorescence enhancement originates from enhanced *Purcell·QE* at the emission wavelength, and is not promoted via enhanced excitation of NV centers in close proximity of gold nanorod.

The *Purcell·QE* product is slightly decreased by 0.63- and 0.95-times, the larger quantum efficiencies indicate considerably less amount of light is absorbed both at the excitation and the emission wavelengths inside the gold nanorod in the simultaneously optimized configuration, than in corresponding (i) and (ii) cases. However, the simultaneously optimized configuration possesses smaller Purcell factors at both wavelengths (Online Resource, Table 1, 2, bottom, left).

In silver nanorod based configurations, three orders of magnitude larger (*Purcell·QE*)$^2$ quantity is achievable. In the configuration exhibiting the highest 25622.65 (*Purcell·QE*)$^2$ quantity 72.42 and 353.79 *Purcell·QE* products are reached, when 1005.55 and 915.31 *Purcell* factors are taken on at the excitation and emission wavelength. These *Purcell* factors are accompanied by 7.20% and 38.65% *QE* values, which show that the non-radiative loss is significantly smaller at the emission wavelength, in accordance with the wavelength dependency of material limits to optical responses (Online Resource, Table 1, 2, middle, left) [13, 16, 22, 57]. Both *Purcell·QE* products are significantly larger than unity at 532 nm / 650 nm, which shows large radiative rate enhancement. The *Purcell* factor and *QE* shows a local and a global maximum at the 650 nm wavelength of emission similarly to the case of gold nanorod based optimal configuration determined via (iii) optimization. The advantage of silver is that both quantities have enhanced values at 532 nm, since the *Purcell* factor and *QE* exhibits local maxima close to the wavelength of excitation (Fig. 6a). The wavelength dependency of the *Purcell·QE* product shows a global maximum exactly at 650 nm, and a well-defined local maximum appears very close to 532 nm as well (Fig. 6b). These extrema prove that the net emission enhancement originates from enhanced *Purcell* and *QE* both at the emission and excitation wavelength of NV centers in close proximity of silver nanorod. The enhanced *Purcell* factors are almost equal at the two wavelengths, while the quantum efficiency is five-times better at the wavelength of emission.

Comparing the *Purcell·QE* factors determined by (iii) dual wavelength optimization to those reached via (i) and (ii) independent optimizations we can conclude, that slight 1.11-/0.81-times larger/smaller radiative rate enhancement is achievable at the excitation/ emission wavelength via silver nanorod. The slight increase correlates with the increased *QE* at the excitation wavelength, while the slight decrease occurs in spite of the one order of magnitude larger *QE* at the wavelength of emission, and is caused by the smaller *Purcell* factor achievable in the configuration optimized for two wavelengths (Online Resource, Table 1, 2 bottom, left).

The *Purcell·QE* is 120.28-times and 39.67-times better at the excitation and emission wavelength in case of silver nanorod based configuration optimized to enhance the NV center emission, compared to the corresponding gold nanorod-NV center configuration. The larger *Purcell* factors as well as the larger *QE* at the excitation and emission wavelength with respect to gold indicate that larger total decay rate enhancement and reduced non-radiative losses play similar role. The relative increase of both quantities is larger at the excitation wavelength indicating that this phenomenon is more strongly promoted in case of silver with respect to gold. The radiative rate enhancement wavelength dependency is in accordance with that 532 nm and 650 nm is close to / very far from the interband transitions in case of gold/silver (Online Resource, Table 1, 2 middle, left) [16, 57]. Moreover, the differences in material related limits to optical responses gradually decrease in the 532-650 nm interval in between gold and silver [13, 22].

### 3.2.2 Gold and silver nanorod configurations optimized for excitation and emission wavelength of SiV
As a result of optimization performed to enhance excitation and emission of SiV centers 21.83-times higher (*Purcell·QE*)$^2$ quantity was reached in case of gold nanorod based configuration, than in case of NV center. The highest 117.23 value originates from 1.04 and 112.93 *Purcell·QE* products, which are reached, when 1848.31 and 949.02 *Purcell* factors are taken on. These large *Purcell* factors are accompanied by 0.06% and 11.90% *QE* values, showing that the non-radiative loss is by two orders of magnitude smaller at the wavelength of emission (Online Resource, Table 1, 2 middle, right). The *Purcell·QE* product is slightly/significantly larger than unity at 532 nm / 738 nm, which shows that small/large radiative rate enhancement is achievable at the excitation / emission via gold nanorod.

In case of the dipole corresponding to the excitation, there is a global maximum at a wavelength in between 532 nm and 738 nm both on the *Purcell* factor and *QE*. As a result, there is a corresponding global maximum on the *Purcell·QE* product in this wavelength interval (Fig. 7a, b). In case of the dipole corresponding to the emission, a local and a global maximum appears at 738 nm on the *Purcell* factor and *QE*, however, neither of these optical responses show a maximum at 532 nm, similarly to the configuration optimized for NV center enhancement (Fig. 7a). On the contrary, the *Purcell* factor and *QE* exhibit a global and a local maximum in between the excitation and emission wavelengths.

As a result, the wavelength dependent *Purcell·QE* product shows the global maximum exactly at 738 nm, but do not show extremum at 532 nm, while exhibits a local maximum in between these two wavelengths (Fig. 7a, b). These tendencies indicate that the fluorescence enhancement mainly originates from enhanced *Purcell* and *QE* at the emission wavelength, but is slightly promoted via enhanced excitation in proximity of gold nanorod in case of SiV centers, which is mainly caused by the material related limits to optical responses [13, 16, 22, 57].

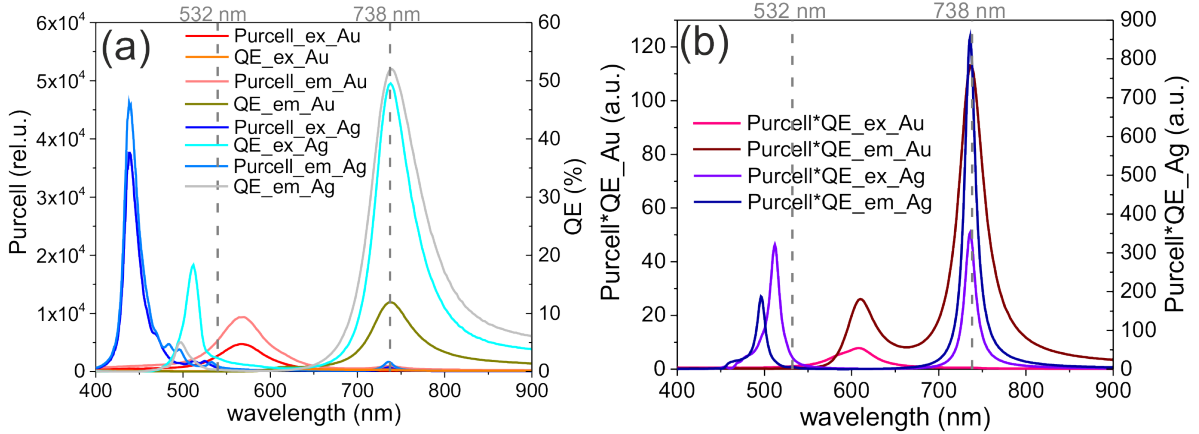

**Fig. 7** (a) Wavelength dependency of the (a) *Purcell* factor and *QE* quantum efficiency, (b) *Purcell·QE* product, in configurations based on gold and silver nanorod, optimized for excitation and emission wavelength of SiV center

The *Purcell·QE* quantities are 1.08- and 2.56-times larger, and the larger *QE*s indicate that slightly and significantly less amount of light is absorbed at the excitation and the emission wavelengths inside the gold nanorod in the configuration (iii) optimized simultaneously, than in corresponding (i) and (ii) cases. The *Purcell* factors are decreased, but with smaller extent compared to the simultaneous optimization performed for NV center (Online Resource, Table 1, 2 bottom, right).

The *Purcell·QE* is 1.72-times and 12.66-times larger at the wavelength of excitation and emission, respectively, than in case of (iii) optimization performed for NV. The small relative increase at the wavelength of excitation is caused by the one order of magnitude decrease in *QE*, which is caused by the difference in the location and orientation of the dipole corresponding to the excitation (Online Resource, Table 1, 2 bottom, middle). The large relative increase at the wavelength of emission is in accordance with the larger distance from interband transitions, and the wavelength dependency of the material related limits to optical responses [13, 16, 22, 57].

In silver nanorod based configuration slightly smaller (*Purcell·QE*)$^2$ quantity is achievable compared to NV center. In the configuration exhibiting the highest 22099.52 (*Purcell·QE*)$^2$ quantity 31.05 and 711.71 *Purcell·QE* products are reached, when 1001.20 and 1347.81 *Purcell* factors are taken on. These *Purcell* factors are accompanied by 3.10% and 52.80% *QE* values at the excitation and emission wavelength, which show a non-radiative loss at the emission smaller than the limit for optimal per-volume scattering predicted theoretically based on plane wave illumination of silver spheroids for the given wavelength (Online Resource, Table 1, 2, middle, right) [13, 22]. The *Purcell·QE* product is considerably/significantly larger than unity at 532 nm / 738 nm, which shows that large/significant radiative rate enhancement is achievable at the excitation / emission via silver nanorod.

In case of silver nanorod the dipoles corresponding to the excitation and emission show very similar optical responses. The *Purcell* factors and *QE* quantities show a local and a global maximum at 738 nm for both dipoles, similarly to the case of gold nanorod based optimal configuration. In addition to this, global and local maxima appear on the *Purcell* factor and on the *QE* quantities at wavelengths significantly and slightly smaller than 532 nm (Fig. 7a). As a consequence, the wavelength dependency of the *Purcell·QE* products shows well defined smaller/larger global maximum exactly at 738 nm for both dipoles, while larger/smaller local maximum appears at a wavelength slightly smaller than 532 nm for the dipole corresponding to excitation/emission (Fig. 7b). The enhanced *Purcell* factors are similar at the two wavelengths, while the quantum efficiency is significantly better at the emission. The advantage of silver is not completely fruited at 532 nm, which can be caused by the perpendicularity of the dipoles corresponding to the excitation and emission.

Comparing the *Purcell·QE* products reached in (i), (ii) and (iii) optimizations we can conclude, that the radiation rate enhancement excitation and emission is 0.48- and 0.46-times less enhanced in (iii) case. The lower excitation enhancement originates from the loss, which causes *QE* decrease with the same extent at the excitation wavelength.

This is the only configuration, where the *QE* is decreased at the excitation in (iii) case, which is most probably caused by the forced relative orientation of the corresponding dipole. The lower emission can be explained with that even though the *QE* is one order of magnitude larger, the *Purcell* factor is more strongly decreased at the wavelength of emission (Online Resource, Table 1, 2 bottom, right). The most significant relative decrease in *Purcell·QE* is in accordance with the peculiarity of the (iii) optimal configuration regarding the significantly enhanced/decreased *QE* / *Purcell* factor.

As a consequence, the *Purcell·QE* is 0.43-times smaller at the wavelength of excitation, while it is 2.01-times larger at the wavelength of emission, than in case of (iii) optimization performed for NV. These differences are mainly caused by the relatively decreased/increased *QE*, while the *Purcell* factor is unaltered/increased at the wavelength of excitation/emission. The former indicate that it is tricky to achieve *QE* enhancement at the excitation wavelength with a forced dipole orientation corresponding to excitation even in close proximity of silver nanorod. The latter is in accordance with the wavelength dependency of material related limits to optical responses, which result in that *Purcell·QE* radiative rate enhancement can be larger at 738 nm than at 650 nm (Online Resource, Table 1, 2, bottom, middle) [13, 16, 22, 57].

In spite of the mismatch between extrema on the optical responses and the excitation wavelength, the *Purcell·QE* is 29.91-times and 6.30-times better at the excitation and emission wavelength in case of the silver nanorod based configuration than in gold nanorod based configuration. The smaller/larger *Purcell* factors with respect to gold indicate that the total decay rate enhancement plays less/more significant role, when SiV center fluorescence is enhanced via silver nanorod at 532 nm / 738 nm. In contrast, the *QE* enhancements with respect to gold prove that the non-radiative loss is significantly/considerably smaller in silver nanorod based configuration at the excitation/emission wavelength (Online Resource, Table 1, 2, middle, right).

The anomalous behavior manifesting itself in smaller *QE* at the same excitation wavelength in case of SiV than in NV in both cases of gold and silver nanorod is caused by that the optimal configuration is determined by the geometry corresponding the emission wavelength (see in section 3.2.4). This proves that the maximal improvement achievable at the wavelength of excitation depends on the excitation-emission wavelength combinations according to Eq. (2), when optimization is performed for two wavelengths simultaneously. These results indicate that configuration optimization is especially a challenge, when the dipoles corresponding to excitation and emission are perpendicular to each other, as in case of SiV (Online Resource, Table 1, 2 bottom, middle).

Most important result of present work is that the apparent quantum efficiency of the coupled system is larger than the intrinsic (~10%) quantum efficiency of SiV (Online Resource, Table 2, middle, right), namely the 11.82 and 52.54 enhanced quantum efficiencies correspond to 1.18-times and 5.25-times enhancements. These results prove that it is possible to enhance the quantum efficiency of weak color centers in diamond via appropriately designed nanorod based coupled systems.

### 3.2.3 Comparative study of near-field and charge distributions in configurations optimized for excitation and emission wavelengths

The inspection of the time averaged power flow, **E**-field distribution and the accompanying charge distribution in the optimized configurations uncover the plasmonic modes that are at play at the extrema and help to understand the limits in the achievable (*Purcell·QE*)$^2$ quantity.

In case of configurations optimized to enhance NV center fluorescence, the geometry of gold nanorod is almost spherical, namely the aspect ratio is 1.10, while more elongated silver nanorod with 1.49 aspect ratio is the optimal one. The dipole is located at 6.58 nm from the gold nanorod and exhibits 24.57° inclination, while the dipole distance is 2.33 nm from the silver nanorod and is accompanied by -1.61° tilting. Important difference is that localized quadrupolar surface mode appears at 532 nm in case of gold, while quadrupolar volume mode develops on silver. As a result, the **E**-field is enhanced locally in proximity of the point dipole and the power flow is trapped at the gold nanorod, while on the silver nanorod quadrupolar cloverleaf **E**-field enhancement is observable, which is accompanied by power flow vortices emanating from the coupled system towards four directions. At 650 nm the charge distribution is dipolar on both nanorods, however larger charge separation occurs on silver due to the larger aspect ratio. Accordingly, larger dipolar **E**-field enhancement is observable around the silver nanorod, which is accompanied by stronger power outflow. The directivity of the power flow is determined mainly by the orientation of the dipolar emitter. These coupled configurations promote mainly the emission of the NV center in proximity both of gold and silver (Online Resource, Table 1, 2 middle, left).

In case of configurations optimized for SiV center elongated gold and silver nanorods with 1.54 and 1.88 aspect ratio are optimal. Localized dipolar surface and monopolar modes appear at 532 nm in case of gold and silver, respectively. As a result, the **E**-field is locally enhanced in proximity of the point dipole and the power flow is trapped at the gold nanorod, while **E**-field enhancement is observable on the complete perimeter of the silver nanorod, which is accompanied by power flow vortices surrounding the coupled system, mainly concentrated at the dipole side. At 738 nm the charge distribution is dipolar on both nanorods, however, again larger charge separation occurs on silver. Accordingly, larger dipolar **E**-field enhancement is observable around the silver nanorod, which is accompanied by stronger power outflow. The directivity of the power flow is determined again mainly by the orientation of the dipolar emitter.

The dipoles are at 2.94 nm distance from the gold nanorod and the intermediate 40.2° and -49.8° inclinations corresponding to the dipole oscillating at 532 nm and 738 nm compromise between the enhancements of excitation and emission in case of gold. The dipoles are closer at 2.16 nm distance from the silver nanorod, however, the -11.01° and 78.99° inclinations show that the dipole corresponding to excitation/emission is significantly less/more tilted with respect to the surface normal of the silver nanorod. Complementary studies revealed, that by interchanging them smaller (*Purcell·QE*)$^2$ quantity is achievable (Online Resource, Table 1, 2 middle, right).

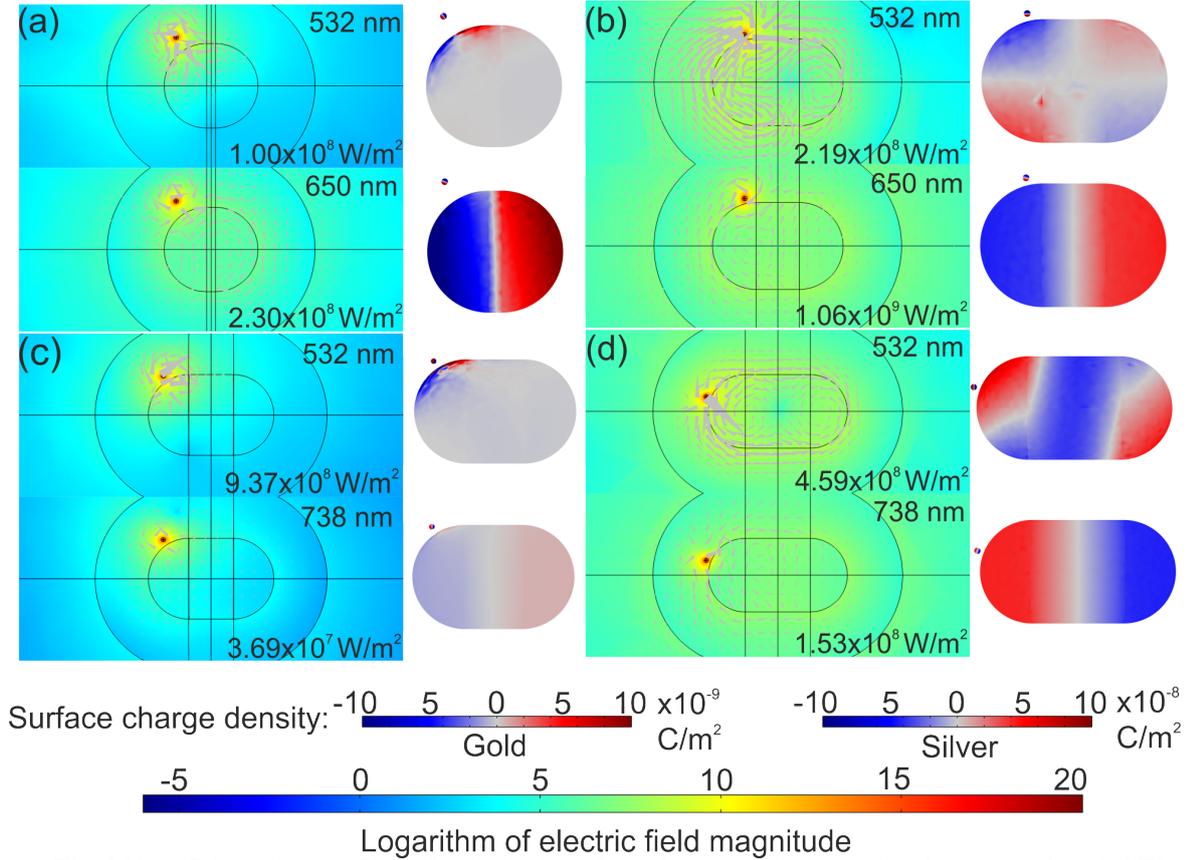

**Fig. 8** Near-field and power flow of optimized configurations at excitation and emission wavelength. NV center near (a) gold and (b) silver nanorod, SiV center near (c) gold and (d) silver nanorod. Insets: corresponding charge density plots at excitation and emission wavelengths. Note that the arrows indicate the total power flow including the radiated and non-radiated components in the coupled system.

Comparing the configurations optimized for single wavelengths and for dual wavelengths of excitation and emission one can conclude, that in the latter case the aspect ratio of the nanorods is determined by the optimal aspect ratio corresponding to the emission. Moreover, the distances of the dipoles from nanorods in configurations optimized for the excitation and emission wavelength simultaneously are more likely to the distances determined by optimization performed for the single wavelength of emission (Online Resource, Table 1, 2 bottom).

In diamond environment having large refractive index, the quadrupolar volume modes can appear already at 532 nm on silver nanorods, while pure dipolar modes are excitable at 650 nm and 738 nm with high efficiency on both nanorods [16]. The quadrupolar volume charge distribution and the large dipolar charge separation are accompanied by power flow rearrangement, which promotes *QE* improvement by decreasing the amount of power lost in the nanorod [16, 17]. Important advantage of quadrupolar modes is that they are capable of resulting in larger radiative decay rate enhancement and are accompanied by narrow linewidth corresponding to higher resonance's quality factor [16]. These phenomena ensure improved *Purcell·QE* at 532 nm as well as at 650 nm excitation and emission wavelength of NV center in case of silver, while in case of gold the excitation is not promoted at 532 nm caused by the localized quadrupolar surface charge distribution.

In contrast, in case of SiV center the dipolar localized surface charge distribution and the monopolar volume charge distribution can only moderately promote the *Purcell·QE* improvement at the wavelength of excitation in close proximity of gold and silver nanorod. As a consequence, the *Purcell·QE* radiative rate enhancement is slightly larger/smaller at the excitation in case of gold/silver nanorod, while it is larger at the emission for both metals, than in case of NV. Further studies are in progress to find configurations capable of better enhancing both excitation and emission in case of SiV centers in diamond.

## 4. Conclusion

Optimization of configurations based on gold and silver nanorods was realized to enhance excitation and emission or both phenomena in case of NV and SiV color centers in diamond. The robustness of the GLOBAL optimization algorithm has been demonstrated via optimizations performed for single and dual wavelengths of excitation and emission.

Both (i) and (ii) single and the (iii) dual wavelength optimizations revealed that enhancement of excitation at 532 nm is not possible by diamond coated gold nanorods, caused by the closer interband transitions [16, 57], larger fraction of plasmon energy in metal [13] and material limits related to optical responses in case of gold [22], while silver nanorods are capable of improving the excitation as well. The material limits allow to reach better *Purcell·QE* products, as a result larger radiative rate enhancements are reached at both wavelengths via silver nanorod, therefore silver is proposed for fluorescence enhancement of the color centers in diamond.

Single wavelength optimizations proved that both metals are capable of enhancing the fluorescent light emission at 650 nm and 738 nm, larger *Purcell·QE* is achievable for SiV center emission via both metals, and in silver nanorod based configuration for both color centers. These results are in accordance with the wavelength dependency of material limits to optical responses [13, 16, 22, 57].

In optimizations performed to enhance excitation and emission simultaneously all tendencies and the near-field study proved that the optimal configurations are determined by the optimal parameters corresponding to the wavelength of emission. Accordingly, the aspect ratios are almost equal and the dipole positions are very similar in corresponding configurations determined by (ii) and (iii) optimization. The *Purcell·QE* is larger in case of SiV at the emission wavelength for both metals, while in silver nanorod based configurations at the excitation wavelength it is decreased along with the *QE* with respect to the NV center. This is caused by perpendicularity of dipoles corresponding the excitation and emission, since in the same configuration the largest *Purcell·QE* is reached along with the largest *QE* via a dipole emitting perpendicularly.

Interestingly the radiative rate enhancements achieved by dual wavelength optimization are slightly different from enhancements resulted by single wavelength optimizations. In SiV - silver nanorod coupled configuration the *Purcell·QE* is decreased with respect to the configuration determined by optimization independently for the single wavelength of excitation and emission. It seems to be contradictory that the *Purcell·QE* at both wavelength is decreased, when in silver nanorod based system the highest (*Purcell·QE*)$^2$ is accompanied by a *QE* overcoming the theoretically predicted material limit for optimal per-volume normalized scattering, therefore further optimizations are still in progress via antenna-like silver nano-objects.

As a result of our work configurations capable of enhancing the fluorescence light emission from NV and SiV centers were determined. Gold and silver nanorod based configurations are presented, which make possible to improve the quantum efficiency to 11.82% and 52.54%, i.e. by 1.18-times and 5.25-times with respect to the intrinsic quantum efficiencies of SiV via coupled plasmonic modes. The proposed configurations are very simple, since nanorods can be prepared by simple colloid chemistry. Further studies are in progress on the optimization of more complex structures.